\documentclass[letter,twocolumn,nofootinbib,showkeys,10pt]{revtex4}
\usepackage[british]{babel}
\usepackage{amsmath}
\usepackage{makeidx}
\usepackage{amsfonts}
\usepackage{dsfont}
\usepackage{graphicx,epstopdf}
\usepackage[ansinew]{inputenc}
\usepackage[usenames,dvipsnames]{pstricks}
\usepackage{subfigure}
\usepackage{pdfpages}
\usepackage{epsfig}
\usepackage{pst-grad}
\usepackage{pst-plot}
\usepackage[colorlinks,hyperindex]{hyperref}
\usepackage{mathrsfs}
\hypersetup{
colorlinks,
citecolor=blue,
linkcolor=black,
urlcolor=black,
}

\DeclareMathOperator{\e}{e}
\DeclareMathOperator{\sech}{sech}
\DeclareMathOperator{\csch}{csch}

\begin{document}

\title{Bounds on topological Abelian string-vortex and string-cigar from
information-entropic measure }
\author{R. A. C. Correa}
\email{rafael.couceiro@ufabc.edu.br}
\affiliation{  CCNH, Universidade Federal do ABC (UFABC), 09210-580, Santo André, SP, Brazil}
\author{D. M. Dantas}
\email{davi@fisica.ufc.br }
\affiliation{Universidade Federal do Ceará (UFC), 60455-760, Fortaleza, CE, Brazil}
\author{C. A. S. Almeida}
\email{carlos@fisica.ufc.br }
\affiliation{Universidade Federal do Ceará (UFC), Departamento de Física, 60455-760,
Fortaleza, CE, Brazil}
\author{Rold\~ao da Rocha}
\email{roldao.rocha@ufabc.edu.br}
\affiliation{Centro de Matem\'atica, Computa\c c\~ao e Cogni\c c\~ao, Universidade
Federal do ABC (UFABC), 09210-580, Santo André, SP, Brazil}

\begin{abstract}
In this work we obtain bounds on the topological Abelian string-vortex and on the string-cigar, by 
using a new measure of configurational complexity, known as configurational
entropy. In this way, the information-theoretical measure of six-dimensional
braneworlds scenarios are capable to probe situations where the
parameters responsible for the brane thickness are arbitrary. The
so-called configurational entropy (CE) selects the best value of the
parameter in the model. {This is accomplished by minimizing the
CE, namely, by selecting the most appropriate
parameters in the model that correspond to the most organized system, based
upon the Shannon information theory.} This information-theoretical measure of
 complexity provides a complementary perspective to situations where
strictly energy-based arguments are inconclusive. We show  that the higher
 the energy the higher the CE, what shows an
important correlation between the energy of the a localized field
configuration and its associated entropic measure.
\end{abstract}

\maketitle

\section{INTRODUCTION}

In 1948, in a seminal work, Shannon \cite{shannon} introduced the  information theory, whose main goal was to introduce the
concepts of entropy and mutual information, using the communication theory. Therein, the entropy was defined to be a measure of \textquotedblleft
randomness\textquotedblright\ of a random phenomenon. Thus, if a little deal
of information concerning  a random variable is received, the uncertainty
decreases, which makes it possible to measure the decrement in the uncertainty, related to the quantity of transmitted information. 
 Inspired by Shannon, Gleiser and Stamatopoulos (GS) latterly introduced a
measure of complexity of a localized mathematical function \cite%
{PLBgleiser-stamatopoulos}. GS proposed that the Fourier modes of
square-integrable, bounded, mathematical functions can be used to construct a
measure, the so-called configurational entropy (CE). A single mode system has zero CE, whereas 
that one where all modes contribute with equal weight has maximal CE. In
order to apply such ideas to physical models, GS used the energy density of
a given spatially-localized field configuration, as a solution of
the related partial differential equation (PDE). Hence the
CE can be used to choose the best fitting trial
function with energy degeneracy.

The CE has been already employed to
acquire the stability bound for compact objects \cite{PLBgleiser-sowinski},
to investigate the non-equilibrium dynamics of spontaneous symmetry breaking 
\cite{PRDgleiser-stamatopoulos}, to study the emergence of localized objects
during inflationary preheating \cite{PRDgleiser-graham} and to discern
configurations with degenerate-energy spatial profiles \cite%
{Rafael-Dutra-Gleiser}. Moreover, solitons were studied in a Lorentz
symmetry violating (LV) framework with the aid of CE \cite{Rafael-Roldao,
RafaelPRD, AHEP, RafaelPRD2}. In this context, the CE associated
to travelling solitons in LV frameworks plays a prominent role in probing
systems wherein the parameters are somehow arbitrary. Furthermore, the CE identifies 
critical points in continuous phase transitions \cite{stamatopoulos/2015}. Moreover, the
CE can be used to measure the informational organization in the structure of
the system configuration for five-dimensional (5D) thick scenarios. In
particular, the CE plays an important role to decide the most appropriate
intrinsic parameters of sine-Gordon braneworld models \cite{bc}, being further studied both in $f(R)$ \cite{Rafael-Pedro} and $%
f(R,T)$ \cite{Rafael-Pedro2} theories of gravity. In what follows, we present a
brief discussion of 5D braneworld models to treat the CE in six-dimensional
(6D) scenarios.

Randall-Sundrum (RS) models \cite{RS1, RS2} proposed a
warped braneworld scenario, wherein the gauge hierarchy problem is explained
and the gravity zero mode is localized, reproducing  four-dimensional (4D) gravity on the brane. The 5D bulk gravitons provide a small
correction in the Newton law \cite{RS2}. However, this thin model presents
singularities and drawbacks concerning the non-localization of spin 
gauge and fermion fields  \cite{Bajc}. To
solve these problems, some thick models were proposed \cite{5Dthick}.

Soon after the works of RS, an axially symmetric warped 6D
model was proposed by Gergheta-Shaposhnikov \cite{GS1}, called \textit{%
string-like defect} (SD).  
 This scenario further provided the
resolution of the mass hierarchy and a
smaller correction to the Newtonian potential \cite{GS1}, besides the  non-requirement of fine tuning between the bulk cosmological constant and the
brane tension, for the cancellation of the 4D cosmological constant \cite{GS1}%
. Besides, the localization of gauge zero modes is spontaneous even in the
thin brane case \cite%
{Oda1, Oda2}. Fermions fields are trapped through a
minimal coupling with an $U(1)$ gauge background field \cite{Liu1,Liu2}.
Later, other 6D,  spherically symmetric, models were employed to explain the
generations of fundamental fermions \cite{Merab,Aguilar} and the resolution
of the mass hierarchy of neutrinos as well \cite{Frere}. Nevertheless, the SD model is a thin model and it leads to some irregularities
\cite{Tinyakov}. Due to it, some 6D thick models were proposed to solve these remaining issues \cite{Gio,
Koley,Luis,Torrealba:2010sg,Torrealba:2010zz,Silva:2012yj, Diego,
Costa:2015dva, Coni1, Coni2, Davi1, Davi2, Davi3,smoothed,Navarro}. In Ref. \cite%
{Gio}, a topological abelian Higgs vortex was used to construct a regular
scenario in which the dominant energy conditions hold, however solely
numerical solutions have been found. Similarly, Ref. \cite%
{Torrealba:2010sg,Torrealba:2010zz}, looking for an exact vortex solution,
shows that the energy density and the angular pressure are similar. This condition is likewise verified for the Resolved Conifold
scenario \cite{Coni1, Coni2, Davi1}. Finally, 
for the String-Cigar \cite{Silva:2012yj, Diego, Costa:2015dva, Davi3}, the
transverse space is represented by a cigar soliton,  which is a stationary
solution for the Ricci flow \cite{ricci1,ricci2, ricci3}. The dominant energy conditions are also satisfied in this model.

Therefore, in this paper we investigate the entropic measure both in the
Torrealba topological Abelian string (TA) \cite%
{Torrealba:2010sg,Torrealba:2010zz} and String-Cigar (HC) \cite%
{Silva:2012yj, Diego, Costa:2015dva, Davi3} in 6D scenarios  due, its analytic
properties. The main aims of our
work is to find bounds for 6D string defects based upon the CE concept and to establish a value for the thickness of the configuration responsible for extremizing the CE.

This paper is organized as follows: in Sect. II a briefly review
string-like defects is present, whereas in Sect. III the CE bounds the
parameters of TA and HC scenarios. We expose the conclusions and
perspectives in Sect. IV, accordingly.

\section{String-like Defect in Warped Six Dimensions}

%

A metric \emph{ansatz} for 6D string-like models reads \cite{GS1, Oda1} 
\begin{equation}
ds_{6}^{2}=\sigma(r)\eta _{\mu \nu }dx^{\mu }dx^{\nu
}-dr^{2}-\gamma(r)d\theta ^{2}\,  \label{6dmetric}
\end{equation}%
where $\eta _{\mu
\nu }=diag(+1,-1,-1,-1)$. The radial coordinate is
limited to $r\in \left[ 0,\infty \right) $, whereas the angular coordinate
is restricted to $\theta \in \left[ 0,2\pi \right) $. The $\sigma(r)$
represents the dimensionless warp factor and $\gamma(r)$ has length squared
dimension.

The 4D Planck mass ($M_P$) and the bulk
Planck mass ($M_6$) are related through the volume of the transverse of space as \cite{GS1, Silva:2012yj,Diego,Davi3}: 
\begin{equation}  \label{hierarchy}
M^2_P=2\pi M_6^4\int_0^{\infty}{\sigma(r)\sqrt{\gamma(r)}dr} \ .
\end{equation}
In addition, the energy-momentum tensor $T_{M}^{N}=diag\left(
t_{0},t_{0},t_{0},t_{0},t_{r},t_{\theta }\right) $ components are given by 
\cite{GS1, Silva:2012yj} 
\begin{subequations}
\label{tmn}
\begin{eqnarray}
&&\left. t_{0}(r)=-\frac{1}{\kappa}\left( \frac{3\sigma^{\prime \prime }}{%
2\sigma}+\frac{3\sigma^{\prime }\gamma^{\prime }}{4\sigma \gamma}+\frac{%
\gamma^{\prime \prime }}{2\gamma}-\frac{\gamma^{\prime 2}}{4\gamma^{2}}%
\right) -\Lambda ,\right.  \label{t0} \\
&&\left. t_{r}(r)=-\frac{1}{\kappa}\left( \frac{3\sigma^{\prime 2}}{%
2\sigma^{2}}+\frac{\sigma^{\prime }\gamma^{\prime }}{\sigma \gamma}\right)
-\Lambda ,\right. \hspace{1.5cm}  \label{tr} \\
&&\left. t_{\theta }(r)=-\frac{1}{\kappa}\left( \frac{2\sigma^{\prime \prime
}}{\sigma}+\frac{\sigma^{\prime 2}}{2\sigma^{2}}\right) -\Lambda ,\right. 
\hspace{1.5cm}  \label{t6}
\end{eqnarray}%
where the $\kappa=\frac{8\pi}{M_{6}^{4}}$ is the 6D gravitational
constant, $\Lambda$ is the 6D (negative) 
cosmological constant  and the prime denotes the derivative with respect to the radial coordinate 
$r$.

To obtain a regular geometry, the conditions \cite%
{GS1,Gio,Navarro,Silva:2012yj} 
\end{subequations}
\begin{eqnarray}
\sigma(r)\Big|_{r=0} &=&const.,\qquad \sigma^{\prime }(r)\Big|_{r=0}=0,  \notag
\\
\quad \gamma(r)\Big|_{r=0} &=&0,\qquad \left( \sqrt{\gamma(r)}\right)
^{\prime }\Big|_{r=0}=1\,,  \label{regularity}
\end{eqnarray}%
must hold. 

For the vacuum solution, the
warp factor for the Gergheta-Shaposhnikov \textit{String Like Defect} (SD) model is proposed as \cite{GS1,Oda1, Oda2, Liu1, Liu2}: 
\begin{equation}
\sigma_{_{SD}}(r)=\e^{-cr},\quad
\gamma_{_{SD}}(r)=R_{0}^{2}\sigma_{_{SD}}(r)\,  \label{gs-string}
\end{equation}%
where the parameters $c$ is a constant, which
connects the 6D Newtonian constant and the 6D cosmological constant, and $%
R_{0}$ is the radius of compactification of transverse space. See that, in
the limit where $r\rightarrow 0$, only the first condition of Eq. %
\eqref{regularity} holds. 

Following the perspective pointed by Ref. \cite{GS1}, Giovannini in adopted a 6D action \cite{Gio}, wherein the matter Lagrangian is an
Abelian-Higgs model and the transverse space obeys the
Abrikosov-Nielsen-Olesen \emph{ansatz} \cite%
{Gio,Torrealba:2010sg,Torrealba:2010zz}: 
\begin{eqnarray}
\phi (r,\theta ) &=&vf(r)\e^{-il\theta }\qquad l\in \mathds{Z}\ ,
\label{nov}\nonumber \\
\mathcal{A}_{\theta }(r) &=&\frac{1}{q}\left[ l-P(r)\right] ,\quad \mathcal{A%
}_{\mu }=\mathcal{A}_{r}=0\ , \nonumber
\end{eqnarray}%
where $\phi $ and $\mathcal{A}_{M}$ are scalar and gauge fields,
respectively. The condition $v=1$ is a length dimension $L^{-2}$ constant.
The functions $f(r)$ and $P(r)$ are such that $f(r\rightarrow 0)=0$, $%
f(r\rightarrow \infty )=1$, whereas $P(r\rightarrow 0)=l$ and $P(r\rightarrow
\infty )=0$.

From constraints by this \emph{ansatz} and the regular conditions in the Eq. %
\eqref{regularity}, the solutions of fields and warp factors are numerically
obtained in Ref. \cite{Gio}. On the other hand, by imposing conditions on the
function $P(r)\equiv 0$, Torrealba \cite{Torrealba:2010sg,Torrealba:2010zz}
obtained an analytical solution, named \textit{Topological Abelian Higgs string%
} (TA): 
\begin{equation}  \label{ta-string}
\sigma_{_{TA}}(r)=\cosh ^{-2 \delta }{\left( \frac{\beta r}{\delta }\right) }%
,\quad \gamma_{_{TA}}(r)=R_{0}^{2}\sigma_{_{TA}}(r)\ ,
\end{equation}%
where the parameter $\beta$ is similar to the parameter $c$ in the SD model, and $\delta $ is 
a thickness parameter which, for small values, reproduces the thin
Gergheta-Shaposhnikov model in Eq. \eqref{gs-string}. Moreover, Ref. 
\cite{Torrealba:2010sg} concludes that, for the localization of gauge fields
zero mode, the thickness of the model can not exceed the value 
\begin{equation}  \label{delta1}
\delta<\frac{5\beta}{4\pi} q^2v^2\ .
\end{equation}
Now, in the TA \eqref{ta-string} string, two of the conditions (\ref%
{regularity}) are verified.

In another approach, the transverse space can also be built for a cigar
soliton solution of Ricci flow \cite{Silva:2012yj, Diego, Costa:2015dva,
Davi3} 
\begin{equation}
\frac{\partial }{\partial \lambda }g_{MN}(\lambda )=-2R_{MN}(\lambda )\ , \nonumber
\end{equation}%
with $\lambda $ being a metric parameter Ref. \cite{Silva:2012yj, Diego,
Costa:2015dva, Davi3} constructed the geometry named \textit{Hamilton String
Cigar} (HC), where the warp factors read 
\begin{equation}
\!\!\!\!\sigma_{_{HC}}(r)=\e^{-cr+\tanh (cr)}, \;\; \gamma_{_{HC}}(r)=\frac{%
\tanh^2 {cr}}{c^2}\sigma_{_{HC}}(r)\,.  \label{hc-string}
\end{equation}%
In this case, all conditions of Eq. \eqref{regularity} do hold.

\begin{figure}[!htb]
\includegraphics[width=5.8cm]{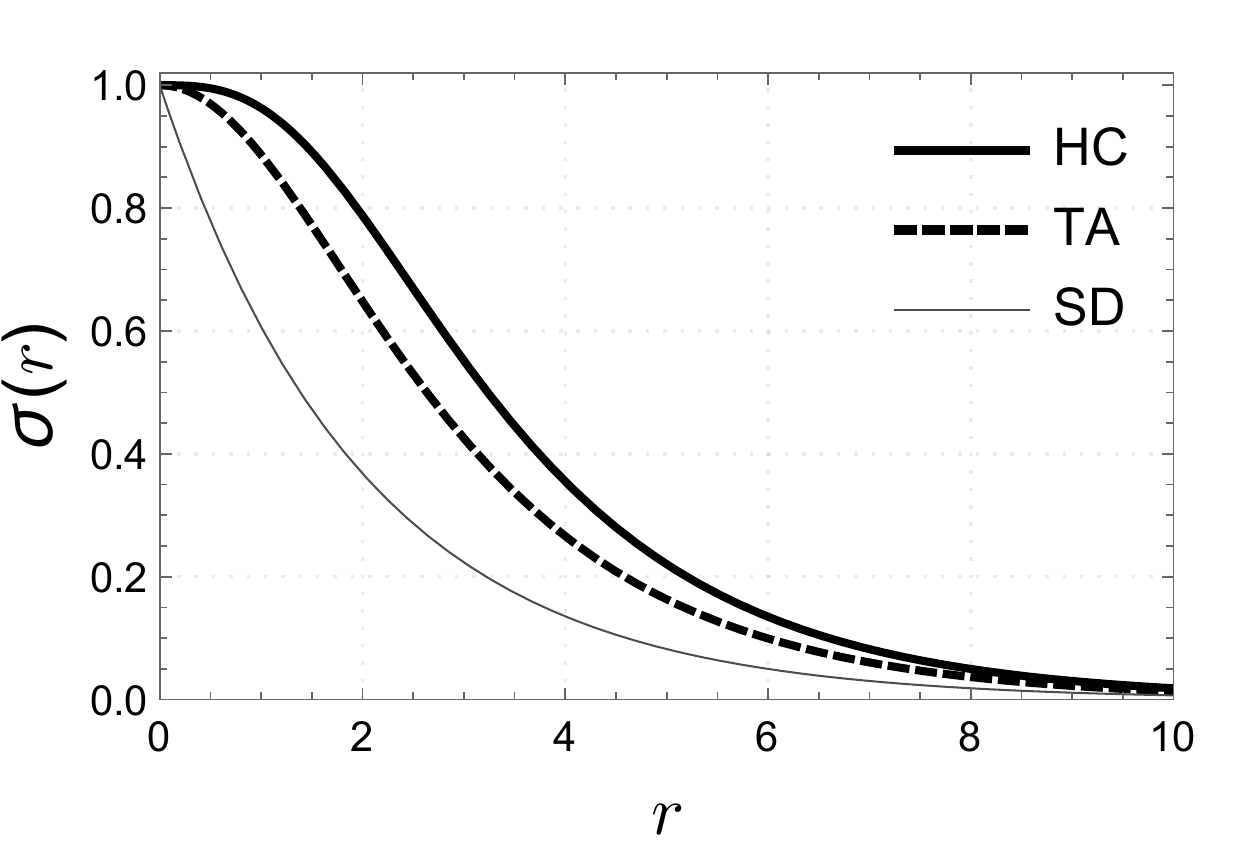}
\caption{$\protect\sigma(r)$ warp-factor with $c = 2\protect\beta= \delta= 0.5$. In the TA (dashed lines) and HC model (thick lines)
the regularity conditions \eqref{regularity} are satisfied for this factor.}
\label{fig-f}
\end{figure}

\begin{figure}[!htb]
\includegraphics[width=5.7cm]{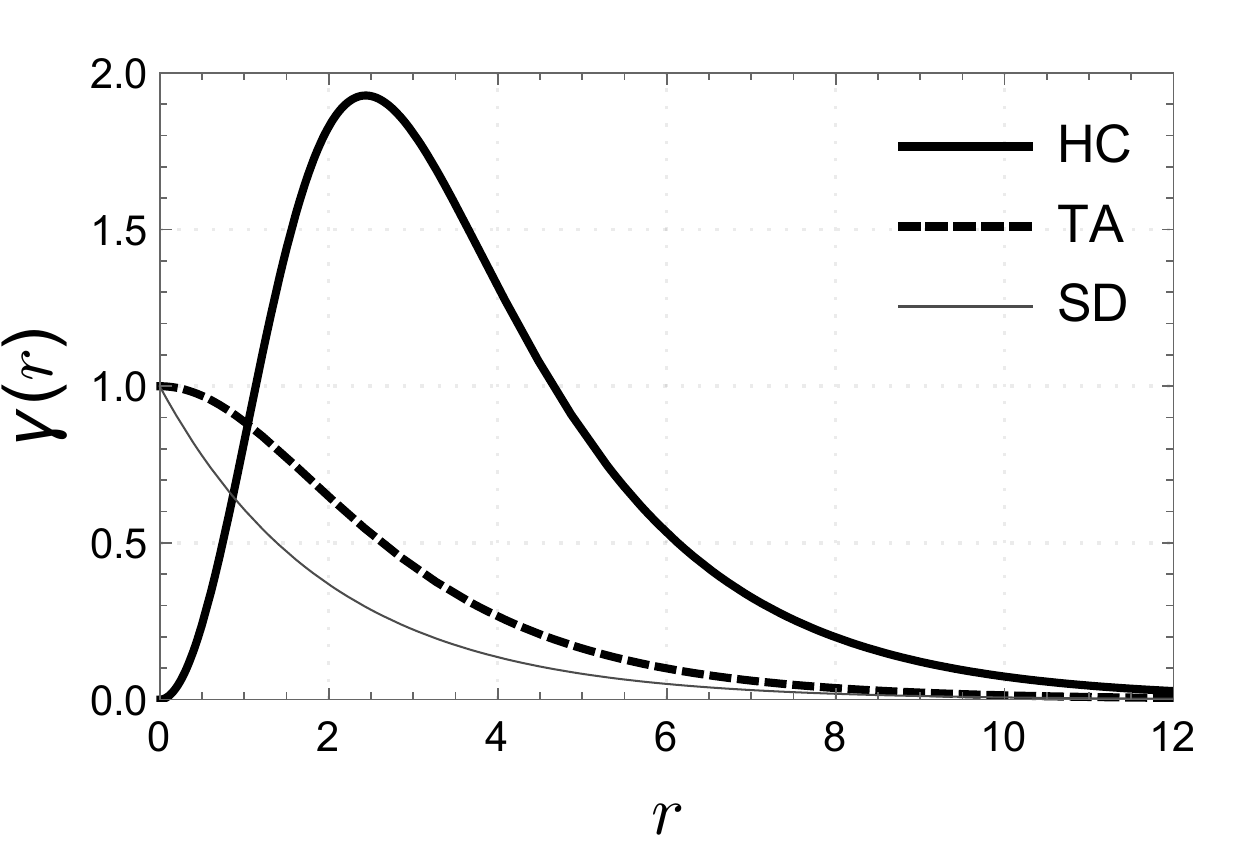}
\caption{$\protect\gamma(r)$ angular factors with $c = 2\protect\beta=\delta= 0.5$ and 
$R_0=1$. Only in the HC model (thick lines) the
regularity conditions \eqref{regularity} hold still.}
\label{fig-h}
\end{figure}

\begin{figure}[!htb]
\includegraphics[width=5.8cm]{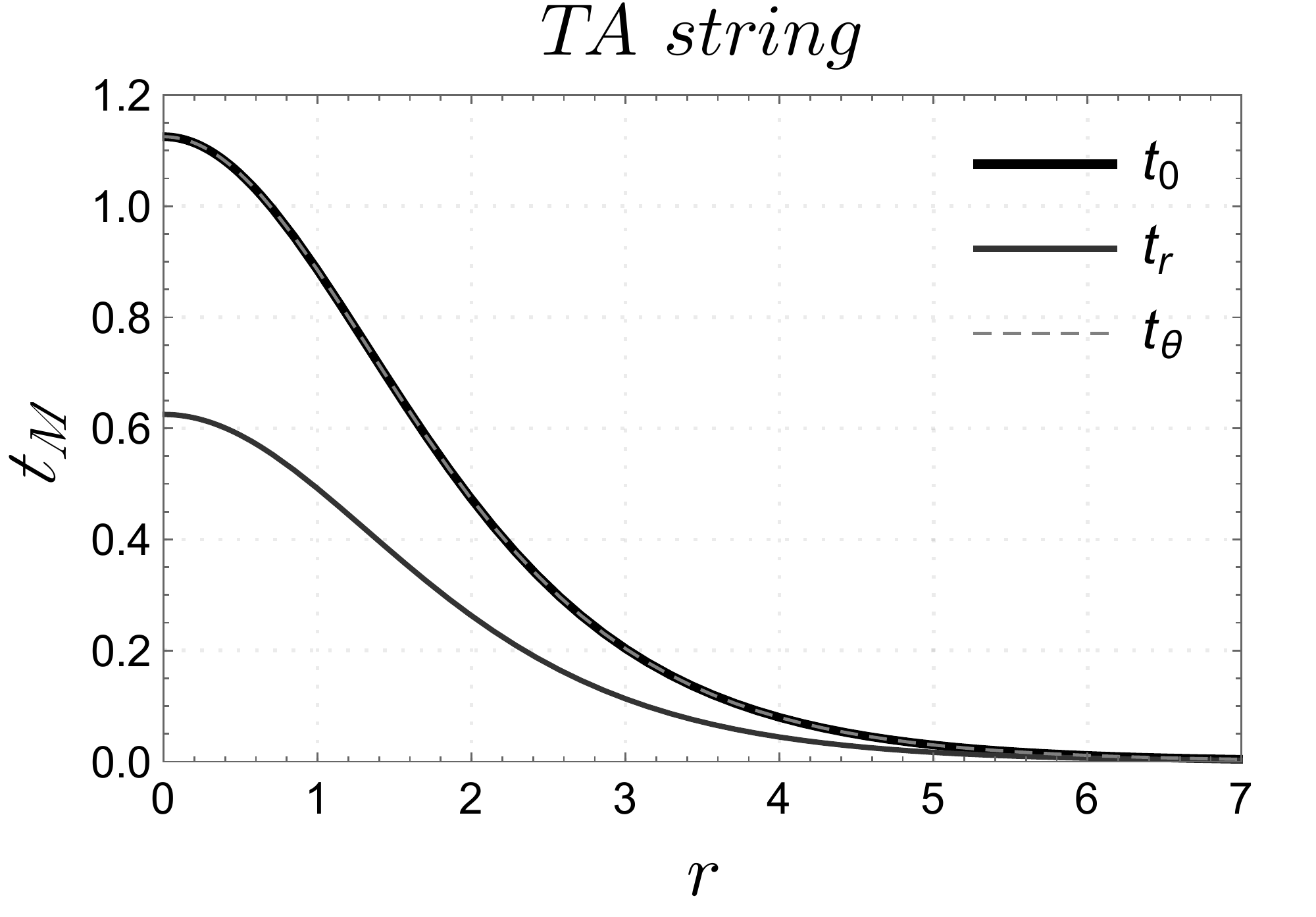} 
\caption{$t_M(r)$ energy-momentum tensor in TA model with $\protect\beta =
0.25$, $\kappa=R_0=1$ and $\protect\delta=0.5$. Here $t_0=t_{\theta}$.}
\label{fig-t-ta}
\end{figure}

\begin{figure}[!htb]
\includegraphics[width=5.8cm]{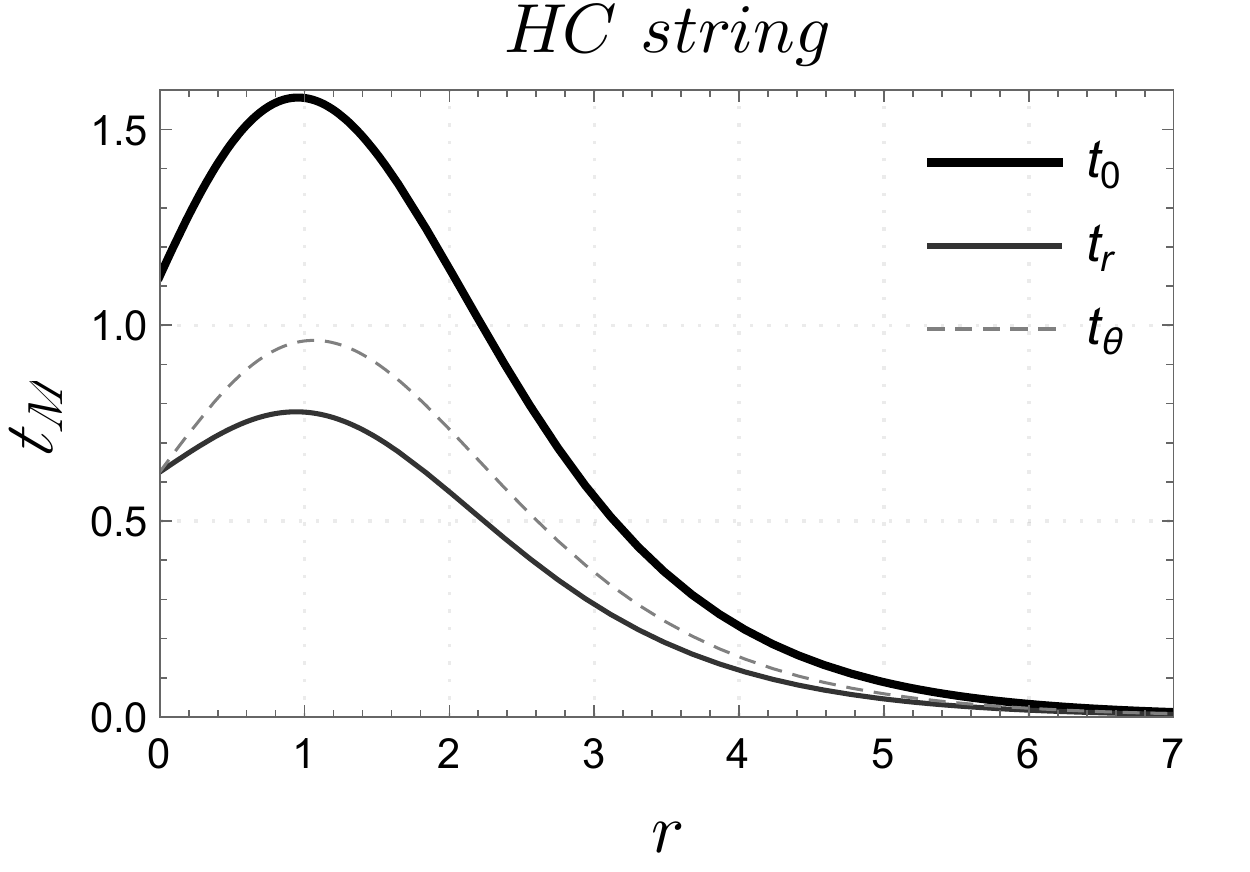}
\caption{$t_M(r)$ in HC model with $c = 0.5$ and $\kappa=1$. The
dominant energy condition is satisfied.}
\label{fig-t-hc}
\end{figure}

To observe the correspondence between the regular condition in Eq. %
\eqref{regularity} and the energy momentum tensor we plot the $\sigma(r)$
the warp factors \eqref{gs-string}, \eqref{ta-string} and \eqref{hc-string}
in Fig. \ref{fig-f} and $\gamma(r)$ in Fig. \ref{fig-h}, whereas the energy
momentum tensor in Fig. \ref{fig-t-ta} for TA and HC in Fig. \ref{fig-t-hc}.
Concerning the HC scenario, wherein all metric conditions \eqref{regularity}
hold, the dominant energy condition $t_{0}\geq \lvert t_{i}\lvert \ , (i=r,
\theta) \,$ \cite{Koley, smoothed,Luis} is satisfied.


In the next section, we shall analyze these string models from the CE
point of view.

\section{Configurational Entropy in the vortex-string scenario}

The configurational
entropy (CE) \cite{PLBgleiser-stamatopoulos} represents an original quantity,
employed to quantify the existence of non-trivial spatially localized
solutions in field configuration space.  The CE is useful to bound the
stability of various self-gravitating astrophysical objects \cite{Gleiser-Jiang}, bound states in LV scenarios \cite%
{Rafael-Roldao}, in compact objects like Q-balls \cite{PLBgleiser-sowinski},
and in modified theories of gravity as well \cite{Rafael-Pedro}.
The CE is linked to the energy of a localized
field configuration, where low energy systems are correlated with small 
entropic measures  \cite{PLBgleiser-stamatopoulos}.

The CE can be obtained \cite{PLBgleiser-stamatopoulos} by the Fourier
transform of the energy density $t_{0}(r)$ \cite{bc, Rafael-Pedro}, yielding $
\mathcal{F}(\omega )=-\frac{1}{\sqrt{2\pi }}\int_{0}^{\infty }{t_{0}(r)\e%
^{i\omega r}dr}.$ It is worth to remark that we will consider structures
with spatially localized, square-integrable, bounded  energy density  functions $t_{0}(r)$.
 The  modal fraction reads \cite{PLBgleiser-stamatopoulos,
PLBgleiser-sowinski, PRDgleiser-stamatopoulos, Rafael-Dutra-Gleiser}  $f(\omega )={\lvert \mathcal{F}(\omega )\lvert ^{2}}/{\int_{0}^{\infty }{%
d\omega \lvert \mathcal{F}(\omega )\lvert ^{2}}}\,.$  
Next, the normalized modal fraction is defined as the ratio of the
normalized Fourier transformed function and its maximum value $f_{max}$, namely, $
\tilde{f}(\omega )={f(\omega )}/{f_{max}}$. 
{{}} A localized and continuous function $\tilde{f}%
(\omega )$ yields the following definition for the CE:  
\begin{equation}
S(\tilde{f})=-\int_{0}^{\infty }{d\omega }\tilde{f}(\omega )\mbox{ln}\left[ 
\tilde{f}(\omega )\right] .  \label{CE}
\end{equation}

Therefore, we use this concept to obtain the CE in the Abelian
string-vortex and the string-cigar contexts. By substituting the warp factor %
\eqref{ta-string} in the energy density given by Eq. \eqref{t0}, it yields 
\begin{equation}
t_{0}(r)=\frac{1}{\kappa}\left( \frac{5}{2}+\frac{1}{\beta }\right) \left[ 2\beta %
\sech{\left(\frac{\beta r}{\delta }\right)}\right] ^{2}.  \label{t0-ta}
\end{equation}%
It represents a localized density of energy, as can be verified in Fig. \ref{fig-t-ta}. {Now, the Fourier transform of \eqref{t0-ta} reads}
\begin{equation}
\mathcal{F}(\omega )=\sqrt{2\pi }\delta \omega (5\delta +2)\csch\left( 
\frac{\pi \delta \omega }{2\beta }\right) ,
\end{equation}%
{which is a localized function having the normalized modal fraction:}
\begin{equation}
\tilde{f}(\omega )=\left[ \frac{\pi \delta \omega }{2\beta }\ \mbox{csch}%
\left( \frac{\pi \delta \omega }{2\beta }\right) \right] ^{2}\,.
\label{f-CE-TA}
\end{equation}%
{For the numerical evaluation of Eq. \eqref{CE}, with the input of Eq. \eqref{f-CE-TA}, it is necessary to explicit here the expression of the parameter $\beta$, as defined in Refs. } \cite{Torrealba:2010sg,Torrealba:2010zz}:
\begin{equation}
\beta=\sqrt{\frac{(-\Lambda)\kappa}{10}}, \quad\text{with} \qquad 0<\beta\leq\frac{1}{2}\ ,
\label{b-range}
\end{equation}%
{Let us remember that $\kappa$ is the 6D gravitational constant, being $\Lambda<0$ the 6D cosmological constant. Besides, this imposition over the range of the parameter $\beta$ is necessary to prevent values larger than Planck mass \cite{Costa:2015dva}.}

\begin{figure}[!htb]
\includegraphics[width=6.5cm]{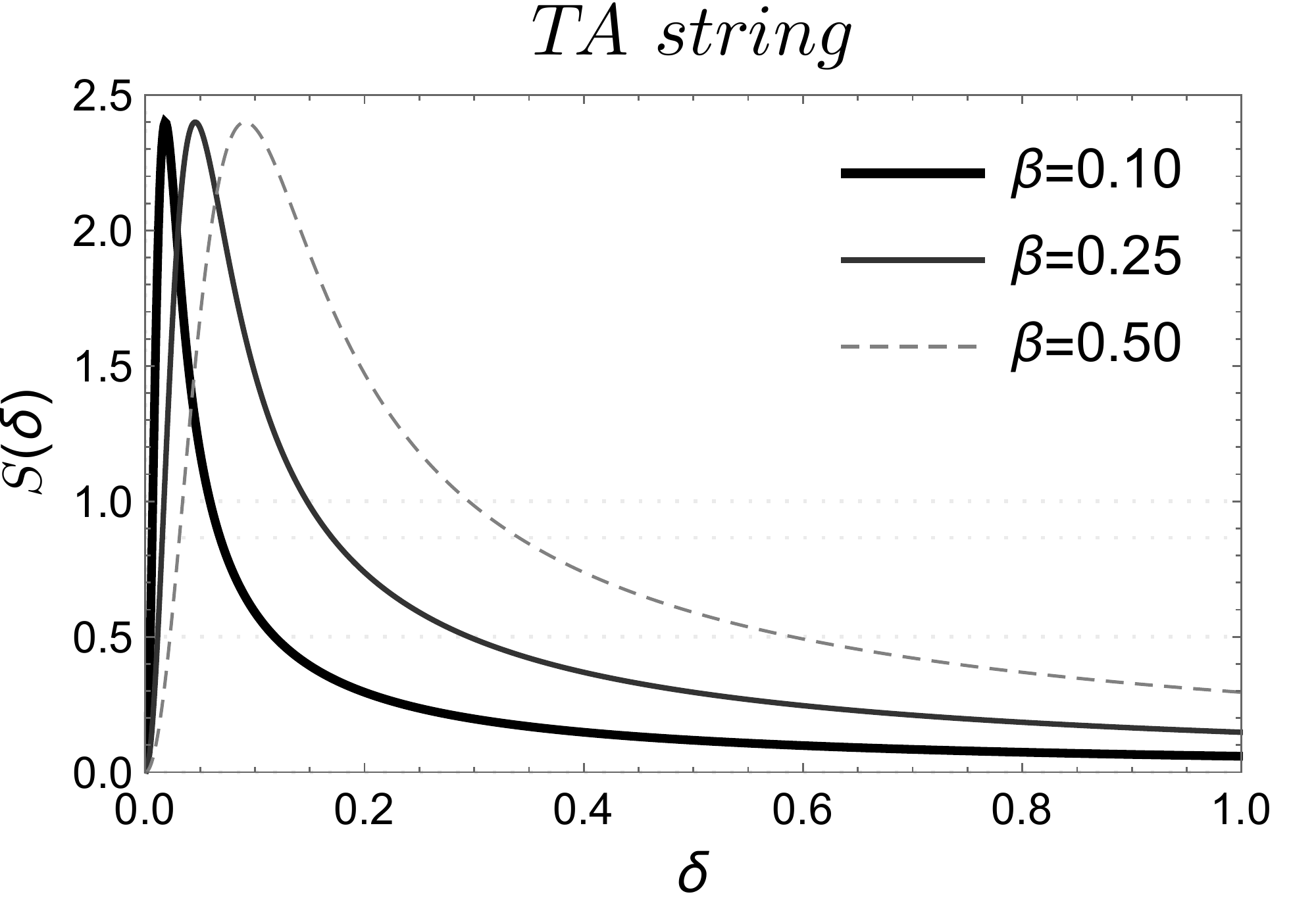}
\caption{$S(\protect\delta)$ Configurational entropy as a function of the thickness parameter $%
\protect\delta$, for different values of
the parameter $\protect\beta$.}
\label{fig-CE-TA}
\end{figure}

\begin{figure}[!htb]
\includegraphics[width=6.5cm]{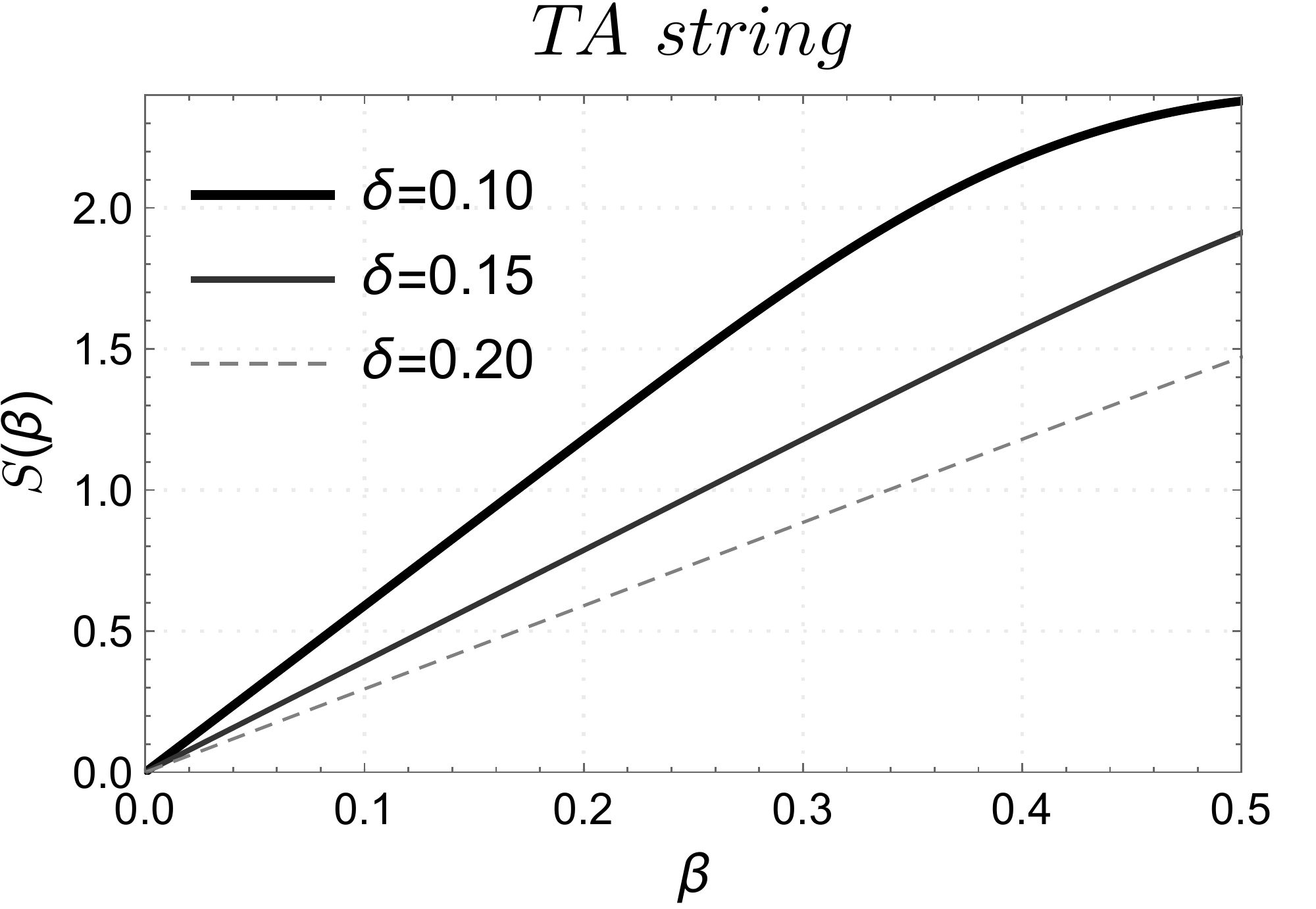}
\caption{$S(\protect\beta)$ Configurational entropy as a function of the parameter $%
\protect\beta$, for different values of
the parameter $\protect\delta$.}
\label{fig-CE-TA2}
\end{figure}

  {Hence the profile of CE in Eq. \eqref{CE} for the function in 
Eq. \eqref{f-CE-TA} is presented in the Fig. \ref{fig-CE-TA}, for $S(\delta)$, and in the Fig. \ref{fig-CE-TA2}, for  $S(\beta)$.} 

{It is verified in Fig. \ref{fig-CE-TA} that the
maximum of CE occurs for $\delta _{crit}\approx 0.09\beta $. This result bounds the thickness of TA model in two regions: the first one for $\delta\to 0$, which endorses the thin Gergheta-Shaposhnikov model of Eq. \eqref{gs-string}, and  the second one for $\delta>\delta _{crit}$. However, an upper bound thickness limit is provided in Eq. \eqref{delta1}.  Thus, for $\delta \neq 0$, the constraint on the TA model thickness  with $q=1$ in the Eq. \eqref{delta1} yields}
\begin{equation}
0.09\beta <\delta <0.40\beta .\,  \label{delta2}
\end{equation}

{Furthermore, another important physical information is presented in Fig. \ref{fig-CE-TA2}, where the minimal CE occurs when the parameter $\beta$ tends to zero. Perceive that the mass hierarchy of Eq. %
\eqref{hierarchy} for the TA model is exposed  in Eq. %
\eqref{ta-string} as}
\begin{eqnarray}
M^2_P=\frac{2\pi R_0}{3}\frac{\sqrt{\pi}}{\beta} \frac{\Gamma \left(\frac{3 \delta }{2}+1\right)}{ \Gamma \left(\frac{3 \delta }{2}+\frac{1}{2}\right)} M_6^4 \,.
\end{eqnarray}%
{In the case where $M_P \gg M_6$, the parameter $\beta$   to tends to zero, once $\delta$ is bounded by Eq. \eqref{delta1}. Thus, the CE exhibits this stable behaviour in Fig. \ref{fig-CE-TA2}  and to small values of $\beta$ there corresponds to  small  values of CE.}

{For the HC model, where we have only the $c$ parameter, the energy density of Eq. \eqref{t0} yields}
\begin{eqnarray}
t_{0}(r)=\frac{c^{2}}{\kappa }\text{sech}^{2}(cr)\left[ 7+\frac{13}{2}%
\tanh (cr) -\frac{5}{2}\text{sech}^{2}(cr)\right] \ .
\end{eqnarray}%

\begin{figure}[tbh]
\includegraphics[width=6.5cm]{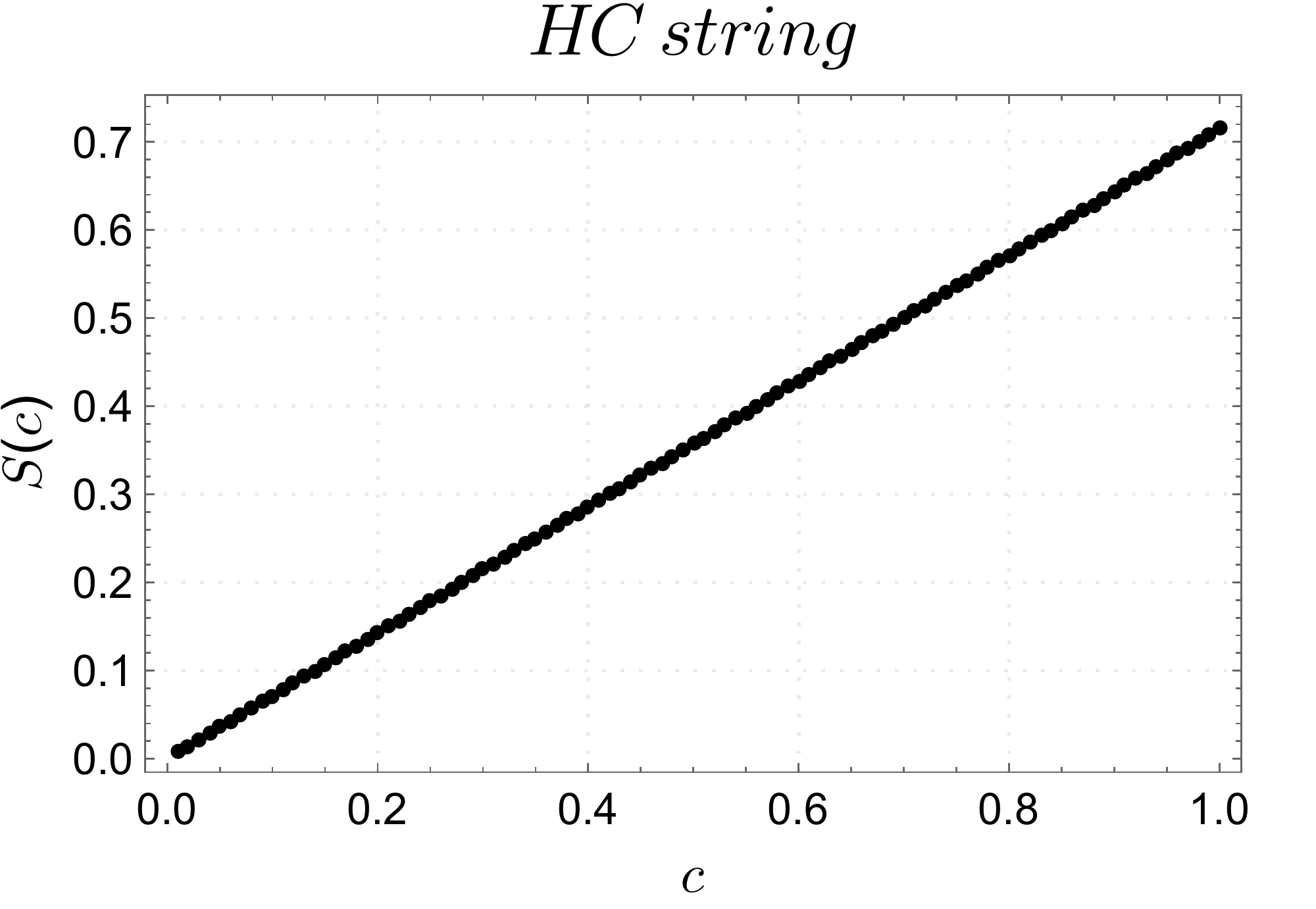}
\caption{$S(\protect\delta )$ Configurational entropy of the HC string
model, as a function of the parameter $c$.}
\label{fig-CE-HC}
\end{figure}

{Again, the energy density is localized as can be verified in the Fig. \ref%
{fig-t-hc}. The Fourier transforms of above equation reads
\begin{eqnarray}
\mathcal{F}(\omega )=\sqrt{\frac{\pi }{2}}\frac{\omega}{12 c^2}  \left(64 c^2+39 i c \omega -5 \omega ^2\right) \csch\left(\frac{\pi  \omega }{2 c}\right) , 
\end{eqnarray}%
and its normalized modal fraction yields}
\begin{equation}
\!\!\!\!\!\tilde{f}(\omega )\!=\!\frac{\pi ^{2}\omega ^{2}\left(
4096c^{4}\!+\!881c^{2}\omega ^{2}\!+\!25\omega ^{4}\right) }{16384c^{6}}%
\csch^{2}\left( \frac{\pi \omega }{2c}\right) .  \label{f-CE-hc}
\end{equation}%
{Setting the expression and the range of the $c$ parameter as defined in Refs. \cite{GS1, Costa:2015dva, Davi3}}
\begin{equation}  \label{c-range}
c=\sqrt{\frac{2\kappa }{5}\left( -\Lambda \right) }, \quad\text{with} \qquad 0<c\leq 1\,
\end{equation}%
{it is possible to plot the $S(c)$ by integrating  Eq. \eqref{f-CE-hc}, using,  \eqref{CE}. Fig. \ref{fig-CE-HC} represents the result. By considering the mass hierarchy of Eq. %
\eqref{hierarchy} in the model of Eq. %
\eqref{hc-string} provided by}
\begin{equation}  \label{gs-hierarchy}
M^2_P\approx\frac{4 \pi R_0}{3}\frac{1}{c} M_6^4 \ ,
\end{equation}%
{the result of $M_P \gg M_6 $ is verified when $c$ tends to zero. This also agrees with the profile exhibited for CE in Fig. \ref{fig-CE-HC}.}

The intrinsic braneworld models parameters have been further constrained by
analyzing the experimental, phenomenological and observational aspects in,
e. g., \cite{bc,epl2012}. In particular, Ref. \cite{bc} provides a refined
analysis wherein the CE further restricts the range parameters of a 5D
sine-Gordon thick braneworld model, namely, the AdS bulk curvature and the
braneworld thickness. {Here this procedure was applied to 6D thick braneworld  models and we verified for the TA model the constraints on the thickness parameter $\delta$ in the Eq. \eqref{delta2}. Besides, in both TA and HC models, the minimal CE reflects the expected result obtained from the mass hierarchy in these models.}

\section{DISCUSSION and CONCLUSIONS}

In this work we have investigated the CE in the context of the topological
abelian string-vortex and string-cigar scenarios. We have shown that the
information-theoretical measure of 6D dimensional braneworld models opens
new possibilities to physically constrain, for example, parameters that are
related to the brane thickness. The CE provides the most appropriate value
of this parameter that is consistent with the best organizational structure. The information measure
regarding the system organization is related to modes related to  the
braneworld model. Hence the constraints of the parameters that we obtained, 
for the TA and the HC string models, provide the range of the parameters
associated to the most organized braneworld models, with respect to the
information content of these models. The CE demonstrates the expected limit of parameters that
agrees with the mass hierarchy of these 6D models. It provides further physical aspects to models, where strictly
energy-based arguments do not provide further conclusions of the physical
parameters.

\section{Acknowledgments}

The authors thank the Coordenação de Aperfeiçoamento de Pessoal de Nível
Superior (CAPES), the Conselho Nacional de Desenvolvimento Científico e
Tecnológico (CNPq), and Fundação Cearense de apoio ao Desenvolvimento Cientí%
fico e Tecnológico (FUNCAP) for financial support. {DMD thanks to Projeto
CNPq UFC-UFABC 304721/2014-0.} RdR is grateful to CNPq grants No.
473326/2013-2, No. 303293/2015-2 and No. 303027/2012-6, and to Fapesp Grant No. 2015/10270-0.
RACC also acknowledges Universidade Federal do Ceará (UFC) for the
hospitality. We would like to thank the anonymous referee for useful suggestions in this letter.

\end{document}